\pdfoutput=1 
\documentclass[preprint,11pt,authoryear]{elsarticle}

\usepackage[utf8]{inputenc}
\usepackage[T1]{fontenc}
\usepackage{lmodern} 
\usepackage{amsmath,amssymb,amsfonts}
\usepackage{graphicx}
\usepackage{hyperref}
\usepackage{natbib}
\usepackage{booktabs}
\usepackage{listings}
\usepackage{xcolor}

\lstdefinelanguage{Julia}{
  keywords={
					function, end, for, while, if, else, elseif, return, using, import, export, module, struct, 
					mutable, const, let, do, try, catch, finally, begin, quote, macro
					},
  sensitive=true,
  morecomment=[l]{\#},
  morestring=[b]",
}
\usepackage{microtype}
\usepackage{multirow}
\usepackage{tabularx}
\usepackage{float}
\usepackage{subcaption}
\usepackage{url}
\usepackage{standalone}
\usepackage{tikz}
\usetikzlibrary{shapes.geometric, arrows.meta, positioning, fit, backgrounds, calc}

\definecolor{excelgreen}{RGB}{00,00,00}
\definecolor{nevenblue}{RGB}{00,00,00}
\definecolor{rcolor}{RGB}{00,00,00}
\definecolor{juliacolor}{RGB}{00,00,00}
\definecolor{pythoncolor}{RGB}{00,00,00}
\definecolor{webviewcolor}{RGB}{00,00,00}
\definecolor{pipecolor}{RGB}{0,0,0}

\usepackage[ruled,vlined,linesnumbered]{algorithm2e}
\usepackage{makecell}
\usepackage{tabularx}
\providecommand{\KwInput}[1]{\textbf{Input:} #1}
\providecommand{\KwOutput}[1]{\textbf{Output:} #1}
\SetKwInput{KwInput}{Input}
\SetKwInput{KwOutput}{Output}
\SetKwComment{Comment}{// }{}

\newcommand{\neven}{NE\raisebox{0.2ex}{<v>}\rotatebox[origin=c]{180}{NE}}

\hypersetup{
    colorlinks=true,
    linkcolor=blue,
    citecolor=blue,
    urlcolor=blue,
    pdftitle={NEVEN: An Open-Source Polyglot Infrastructure},
    pdfauthor={Minor Bonilla-Gómez, Javier Trejos-Zelaya}
}

\journal{}

\begin{document}

\begin{frontmatter}

\title{\large \neven{}: An Open-Source Polyglot Infrastructure Bridging Statistical Computing and Spreadsheet Workflows for Reproducible Research}



\author{
    Minor Bonilla-Gómez \quad Javier Trejos-Zelaya\\[0.5ex]
    {\small\texttt{minor.bonilla@ucr.ac.cr} \quad \texttt{javier.trejos@ucr.ac.cr}}
}


\begin{abstract}
	\footnotesize

Microsoft Excel serves over 750 million users \citep{Microsoft2023}, yet empirical audits indicate that 88\% of operational spreadsheets contain errors \citep{Panko2016}. Although scientific computing environments (R for statistical modeling, Julia for numerical analysis, and Python for machine learning) provide rigorous alternatives, their adoption requires programming proficiency that many domain practitioners lack \citep{Muenchen2012}.

We present \neven{}, an open-source infrastructure that embeds R, Julia, Python, Quarto, and JavaScript visualization libraries directly into Excel through native worksheet formulas. Users invoke econometric estimators, solve differential equations, and compile reproducible reports by entering expressions such as \texttt{=R.MR\_Lineal(...)}, \texttt{=J.Optimize(...)}, or \texttt{=P.ai\_call(...)} in spreadsheet cells without writing procedural code. The platform exposes over 70 R statistical procedures, 70 high-level Julia procedures backed by more than 200 mathematical functions, and full Python 3 integration. 

The system architecture isolates each language runtime in an independent operating system process communicating with the Excel XLL host via Windows Named Pipes and Protocol Buffers serialization. A fatal error in R, Julia, or Python terminates only the corresponding child process while leaving the Excel host operational. A static security sandbox inspects user-submitted code against 25--30 restricted patterns per language (blocking shell execution, file deletion, unauthorized network calls, and dynamic code evaluation) using five bypass-prevention mechanisms validated by 133 dedicated tests.

An embedded WebView2 engine renders interactive \texttt{Plotly} charts, \texttt{D3.js} network and hierarchical visualizations, \texttt{Leaflet} geographic maps, pivot tables, \LaTeX{} mathematical notation, \texttt{PDF} documents, and \texttt{Pluto.jl} reactive notebooks through a unified formula (\texttt{=NEVEN.v()}). In addition, the docked NEVEN Studio environment provides nine specialized modules, including a Chart Gallery with 325 visualization templates, an in-memory DuckDB SQL console, a retrieval-augmented AI assistant, and a multi-language script editor with full English and Spanish localization. Econometric outputs match published benchmarks from \citet{Wooldridge2016} to four decimal places, and linear algebra decompositions in \texttt{Julia} achieve machine-precision accuracy ($<10^{-14}$ relative error). Released under the GNU General Public License v3, \neven{} includes a suite of 342 automated unit, property-based, and integration tests.

\end{abstract}


\end{frontmatter}

\newpage


\section{Introduction}
\label{sec:introduction}

\subsection{The Spreadsheet Dominance Problem}

Microsoft Excel is estimated to serve over 750 million users worldwide \citep{Microsoft2023}, making it the most widely deployed data analysis environment across academic research, corporate finance, public policy, and higher education. Despite this ubiquity, empirical audits have consistently documented that approximately 88\% of operational spreadsheets contain material errors \citep{Panko2016}. High-profile institutional failures illustrate the consequences of fragile spreadsheet workflows: the \$6.2 billion London Whale trading loss was attributed in part to unverified spreadsheet models \citep{JPMorgan2013}; the truncation of public health surveillance records in England resulted from legacy spreadsheet row limits \citep{Kelion2020}; and influential macroeconomic policy conclusions were shown to depend partly on formula range coding mistakes \citep{Herndon2014}.

Within scientific research, concerns over computational reproducibility have been extensively documented \citep{Ioannidis2005, Baker2016}. Spreadsheet-based workflows are particularly susceptible to replication failures because standard workbooks lack version control, programmatic audit trails, and automated testing mechanisms \citep{Peng2011}. Nevertheless, researchers and analysts continue to rely on spreadsheets due to their immediate visual feedback, direct data manipulation model, and low barrier to entry \citep{Scaffidi2005}. These vulnerabilities stem primarily from the absence of a rigorous computational and auditing infrastructure coupled to the spreadsheet interface.

\subsection{The Programming Expertise Barrier}

Scientific computing ecosystems such as R \citep{RCoreTeam2024}, Julia \citep{Bezanson2017}, and Python \citep{VanRossum2009} provide validated, reproducible capabilities across econometrics, numerical optimization, machine learning, and data visualization. However, operating these environments requires programming proficiency that represents a substantial adoption barrier for domain specialists in economics, biomedical sciences, social research, and business analytics. \citet{Muenchen2012} documented that the syntactic learning curve of R remains a primary obstacle among applied researchers. Julia, while delivering native-code numerical performance \citep{Bezanson2017}, presents an equally steep transition for spreadsheet-oriented practitioners.

This divide creates a persistent trade-off between interface accessibility and computational rigor. Existing Excel integration tools, including RExcel \citep{Baier2011}, BERT \citep{BERT2018}, xlwings \citep{xlwings2023}, PyXLL \citep{PyXLL2023}, and Microsoft's Python in Excel \citep{MicrosoftPython2023}, address isolated aspects of the problem. Each of these solutions is restricted to a single programming language and lacks either local process isolation, code sandboxing, or integrated multi-format visualization. Consequently, no unified open-source platform currently bridges multiple scientific runtimes within Excel under a fault-tolerant architecture.

\subsection{Contributions}

This paper presents \neven{}, an open-source polyglot infrastructure designed to bridge spreadsheet workflows and scientific computing through eight primary contributions:

\begin{enumerate}
    \item \textbf{Simultaneous polyglot integration:} A C++17 XLL architecture that unifies R~4.4.1, Julia~1.12.6, Python~3.13 (via the CPython Stable ABI), Quarto, and JavaScript visualization libraries into a single platform accessible through standard Excel formulas. Domain specialists can invoke over 70 R statistical procedures, 70 high-level Julia procedures (backed by more than 200 mathematical functions), and Python analytical routines directly from worksheet cells.

    \item \textbf{Unified multi-format viewer:} An embedded Microsoft WebView2 (Chromium-based) engine that renders interactive Plotly charts, D3.js v7 visualizations, rpivotTable cross-tabulations, Leaflet geographic maps, Markdown documents with \LaTeX{} mathematics, PDF files, Word documents (via Pandoc), and Pluto.jl reactive notebooks through a single worksheet function (\texttt{=NEVEN.v()}).

    \item \textbf{Process-isolated fault-tolerant architecture:} Language runtimes execute in independent operating system processes communicating via Windows Named Pipes with Protocol Buffers serialization. A segmentation fault or unhandled exception in any child runtime never crashes the Excel host. A static security sandbox enforces five bypass-prevention mechanisms across 25--30 restricted patterns per language, verified by 133 automated security tests.

    \item \textbf{Zero-boilerplate user extensibility:} A file-based extension mechanism whereby users register custom R, Julia, or Python functions simply by placing script files (\texttt{.R}, \texttt{.jl}, \texttt{.py}) in a designated user profile directory. Functions are automatically discovered at startup, exposed as native Excel formulas, and documented in the Excel Function Wizard without requiring C++ compilation or administrative privileges.

    \item \textbf{Reproducible scientific publishing:} Native integration with the Quarto publishing system (\texttt{=NEVEN.q()}) to compile publication-ready manuscripts and regulatory reports (HTML, PDF, Word) that combine narrative text, BibTeX citations, and live R, Julia, and Python computations referencing workbook data.

    \item \textbf{Integrated development environment (NEVEN Studio):} A docked Office Web Add-in task pane comprising nine specialized modules, including a Chart Gallery with 325 interactive templates, an in-memory DuckDB SQL console, a retrieval-augmented generation (RAG) AI assistant supporting cloud and local large language models (LLMs), and complete bilingual localization (English and Spanish).

    \item \textbf{Empirical benchmark validation:} Systematic verification of statistical and numerical outputs against the econometric benchmarks of \citet{Wooldridge2016} and matrix factorization identities, confirming that the inter-process serialization pipeline introduces zero numerical degradation.

    \item \textbf{Open-source community architecture:} Released under the GNU General Public License v3 (GPL~v3), \neven{} evolves the single-language R bridge introduced in R4XCL \citep{BonillaR4XCL2026} into a modern, community-extensible polyglot platform backed by 342 automated unit, property-based, and integration tests.
\end{enumerate}

\subsection{Paper Organization}

The remainder of this paper is organized as follows. Section~\ref{sec:architecture} details the four-layer system architecture, initialization sequence, inter-process communication protocol, security sandbox, and visualization engine. Section~\ref{sec:viewer_extensibility} describes the unified viewer, file-based extensibility model, Pluto.jl reactive notebook pipeline, Quarto publishing workflow, and NEVEN Studio environment. Section~\ref{sec:case_studies} presents the \texttt{TipoOutput} extraction paradigm and validates the platform across econometric benchmarks, aerospace trajectory simulation, and enterprise credit risk modeling. Section~\ref{sec:discussion} compares \neven{} against existing tools, reports software quality and security audit metrics, and examines architectural trade-offs and limitations. Section~\ref{sec:conclusion} summarizes the work and outlines four directions for future research.


\section{System Architecture and Design}
\label{sec:architecture}

\begin{figure}[htbp]
    \centering
    \resizebox{0.92\textwidth}{!}{

\definecolor{excelgreen}{RGB}{33,115,70}
\definecolor{nevenblue}{RGB}{41,98,168}
\definecolor{rcolor}{RGB}{39,109,195}
\definecolor{juliacolor}{RGB}{149,88,178}
\definecolor{pythoncolor}{RGB}{210,155,15}
\definecolor{webviewcolor}{RGB}{0,120,212}
\definecolor{pipecolor}{RGB}{90,90,90}

\begin{tikzpicture}[
    every node/.style={font=\sffamily\small},
    layer/.style={draw, rounded corners=2pt, minimum width=14.6cm, minimum height=2.3cm, fill=#1!6, draw=#1!70, thick},
    component/.style={draw, rounded corners=2pt, minimum width=3.0cm, minimum height=1.3cm, fill=#1!14, draw=#1!85, thick, align=center, font=\sffamily\footnotesize},
    pipe/.style={-{Stealth[length=2.2mm, width=1.6mm]}, thick, draw=pipecolor, dashed},
    solidarrow/.style={-{Stealth[length=2.2mm, width=1.6mm]}, thick, draw=#1!85},
]

\node[layer=excelgreen, label={[font=\sffamily\bfseries\footnotesize, text=excelgreen, anchor=south west]north west:Layer 1: Excel Host Interface}] (L1) at (0, 10.2) {};

\node[component=excelgreen] (xll) at (-4.9, 10.2) {\textbf{NEVEN64.xll}\\{\scriptsize C++17 XLL Add-in}};
\node[component=excelgreen] (ribbon) at (0, 10.2) {\textbf{NEVENRibbon.dll}\\{\scriptsize COM Ribbon \& Studio}};
\node[component=excelgreen] (funcs) at (4.9, 10.2) {\textbf{200+ Procedures}\\{\scriptsize \texttt{xlfRegister} API}};

\node[layer=nevenblue, label={[font=\sffamily\bfseries\footnotesize, text=nevenblue, anchor=south west]north west:Layer 2: Core Orchestration \& Security Services}] (L2) at (0, 6.8) {};

\node[component=nevenblue] (config) at (-5.1, 6.8) {\textbf{ConfigService}\\{\scriptsize \texttt{neven-config.json}}};
\node[component=nevenblue] (langmgr) at (-1.7, 6.8) {\textbf{LanguageManager}\\{\scriptsize Process + Pipe Init}};
\node[component=nevenblue] (sandbox) at (1.7, 6.8) {\textbf{SandboxVerifier}\\{\scriptsize 25--30 patterns/lang}};
\node[component=nevenblue] (discovery) at (5.1, 6.8) {\textbf{DiscoveryService}\\{\scriptsize Registry + PATH}};

\node[layer=pipecolor, label={[font=\sffamily\bfseries\footnotesize, text=pipecolor!90!black, anchor=south west]north west:Layer 3: Isolated Language Runtimes (OS Processes)}] (L3) at (0, 3.4) {};

\node[component=rcolor] (controlr) at (-4.9, 3.4) {\textbf{ControlR.exe}\\{\scriptsize R 4.4.1 (70+ procs)}};
\node[component=juliacolor] (controlj) at (0, 3.4) {\textbf{ControlJulia.exe}\\{\scriptsize Julia 1.12.6 (200+ funcs)}};
\node[component=pythoncolor] (controlp) at (4.9, 3.4) {\textbf{ControlPython.exe}\\{\scriptsize Python 3.13 (Stable ABI)}};

\node[layer=webviewcolor, label={[font=\sffamily\bfseries\footnotesize, text=webviewcolor, anchor=south west]north west:Layer 4: Interactive Visualization \& Reproducible Publishing}] (L4) at (0, 0.0) {};

\node[component=webviewcolor] (webview) at (-4.9, 0.0) {\textbf{WebView2 Engine}\\{\scriptsize Plotly, D3.js, Leaflet}};
\node[component=webviewcolor] (pluto) at (0, 0.0) {\textbf{Pluto.jl Server}\\{\scriptsize Reactive Notebooks}};
\node[component=webviewcolor] (quarto) at (4.9, 0.0) {\textbf{Quarto CLI}\\{\scriptsize \texttt{.qmd} $\to$ HTML/PDF/Docx}};

\draw[solidarrow=nevenblue] (xll.south) -- (langmgr.north);
\draw[solidarrow=nevenblue] (xll.south) -- (sandbox.north);

\draw[pipe] (langmgr.south) -- node[pos=0.45, left=2pt, font=\sffamily\scriptsize, text=pipecolor!90!black] {Named Pipes} (controlr.north);
\draw[pipe] (langmgr.south) -- (controlj.north);
\draw[pipe] (langmgr.south) -- node[pos=0.45, right=4pt, font=\sffamily\scriptsize, text=pipecolor!90!black] {Protocol Buffers} (controlp.north);

\draw[solidarrow=webviewcolor] (xll.west) -- ++(-0.85, 0) |- (webview.west);

\end{tikzpicture}}
    \caption{\neven{} four-layer architecture. The C++17 XLL add-in (Layer~1) communicates with isolated language runtimes (Layer~3) via Windows Named Pipes carrying Protocol Buffer messages, orchestrated by Layer~2 services. WebView2, Pluto.jl, and Quarto (Layer~4) provide interactive visualization and reproducible reporting.}
    \label{fig:architecture}
\end{figure}
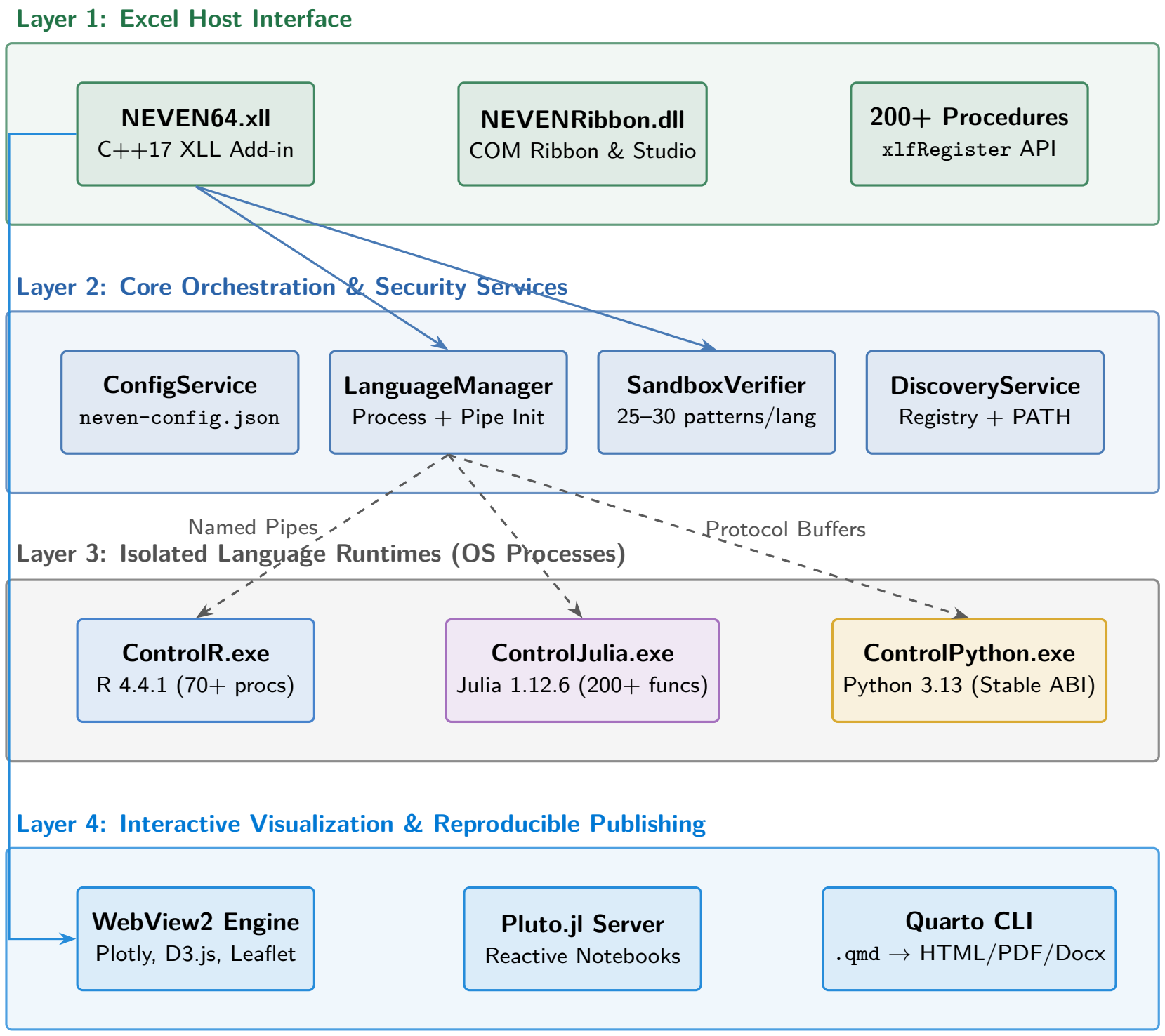

\neven{} is structured as a four-layer architecture (Figure~\ref{fig:architecture}) that decouples the Microsoft Excel host interface from core orchestration services, external language runtimes, and the interactive visualization subsystem. This section details each architectural layer, the inter-process communication (IPC) protocol, and the security verification model.

\subsection{Architectural Overview}

The architectural design enforces three foundational systems engineering principles:

\begin{enumerate}
    \item \textbf{Process fault isolation:} Each language runtime executes in a separate operating system child process. Memory corruption, stack overflows, or segmentation faults within R, Julia, or Python terminate only the affected child process; the Excel host process remains stable and automatically re-establishes the connection upon subsequent calls.
    \item \textbf{Protocol neutrality:} Cross-language data exchange relies exclusively on Protocol Buffers \citep{Protobuf2023}, ensuring language-agnostic binary serialization, strict type safety, and backward-compatible schema evolution.
    \item \textbf{Defense-in-depth code verification:} Dynamic code expressions submitted from worksheet cells or interactive consoles are inspected by a static verification engine and path sanitizer prior to execution.
\end{enumerate}

Historically, \neven{} traces its architectural lineage to the open-source Basic Excel R Toolkit (BERT; \citealp{BERT2018}), which pioneered Named Pipe IPC for R~3.4, and to R4XCL \citep{BonillaR4XCL2026}, the author's master's thesis project at the Universidad de Costa Rica. \neven{} represents a complete C++17 architectural overhaul that expands this paradigm into a simultaneous three-language infrastructure, replaces legacy Electron components with an in-process WebView2 engine, and introduces static sandboxing and automated testing.

\subsection{System Initialization}
\label{subsec:initialization}

The primary binary entry point is \texttt{NEVEN64.xll}, a 64-bit native Excel add-in compiled as a dynamic-link library. When Excel loads the add-in, it invokes the standard \texttt{xlAutoOpen()} callback, triggering the initialization sequence formalized in Algorithm~\ref{alg:init}. During this phase, the XLL dynamically registers approximately 200 built-in and user-defined procedures with the Excel Function Wizard via the C API \texttt{xlfRegister} routine.

\begin{algorithm}[htbp]
\SetAlgoLined
\KwInput{XLL binary loaded by the Microsoft Excel host process}
\KwOutput{System initialized with all language runtimes and formulas registered}
\BlankLine
\texttt{ZombieProcessKiller()}\Comment*[r]{Terminate orphaned child processes}
\texttt{ConfigService::Initialize}(\texttt{"neven-config.json"})\;
\texttt{DiscoveryService::FindLanguages()}\;
\BlankLine
\ForEach{language $L \in \{\text{R}, \text{Julia}, \text{Python}\}$}{
    \If{$L$ is enabled and discovered on host}{
        $proc \leftarrow$ \texttt{CreateProcess}($L$.executable)\;
        $pipe \leftarrow$ \texttt{ConnectNamedPipe}(\texttt{"RJ2XCL2-PIPE-"} $+ L +$ \texttt{"-"} $+$ PID)\;
        \texttt{SendStartupScript}($pipe$, $L$.startup\_file)\;
        \texttt{LoadUserFunctions}($pipe$, $L$.functions\_dir)\;
    }
}
\BlankLine
$functions \leftarrow$ \texttt{MapFunctions()}\Comment*[r]{Query exported descriptors}
\ForEach{$f \in functions$}{
    \texttt{xlfRegister}($f$.name, $f$.params, $f$.description)\;
}
\texttt{ViewerManager::Initialize()}\Comment*[r]{Start WebView2 STA thread}
\texttt{PlutoManager::Initialize()}\;
\caption{System Initialization Sequence (\texttt{xlAutoOpen})}
\label{alg:init}
\end{algorithm}

During the \texttt{LoadUserFunctions} step in Algorithm~\ref{alg:init}, custom scripts placed by the user in \texttt{Documents\textbackslash NEVEN\textbackslash functions\textbackslash} (\texttt{.R}, \texttt{.Jl}, \texttt{.Py}) are parsed by their respective runtimes and registered alongside the core library. Users make new functions available simply by restarting Excel, without requiring C++ compilation, configuration file edits, or elevated system privileges.

The XLL provides two distinct formula dispatch paths:
\begin{itemize}
    \item \textbf{Registered procedure calls} (e.g., \texttt{=R.MR\_Lineal(...)}, \texttt{=J.Algebra(...)}, \texttt{=P.ai\_call(...)}): Routed through \texttt{RJ\_FunctionCall}, which serializes cell arguments into Protocol Buffer messages and dispatches them to the target language process.
    \item \textbf{Ad-hoc code evaluation} (e.g., \texttt{=NEVEN.r("code")}, \texttt{=NEVEN.j("code")}, \texttt{=NEVEN.p("code")}): Routed through \texttt{RJ\_Exec\_Generic}, which subjects the code string to the \texttt{SandboxVerifier} prior to IPC transmission.
\end{itemize}

In addition to formula registration, a companion COM add-in (NEVENRibbon.dll) injects a native Excel Ribbon tab containing 13 action controls organized into five functional groups, providing direct GUI access to language consoles, the unified viewer, Pluto.jl notebooks, Quarto report compilation, and configuration settings.

\subsection{Core Services and Runtime Resilience}

\paragraph{Configuration Service}
The \texttt{ConfigService} parses \texttt{neven-config.json} at startup and exposes strongly typed accessors for all operational parameters. To prevent configuration-based attacks, the parser rejects file paths containing directory traversal sequences (\texttt{..}) or shell metacharacters (\texttt{|}, \texttt{\&}, \texttt{;}), and enforces strict numeric bounds on IPC timeouts and retry counters.

\paragraph{Discovery Service}
The \texttt{DiscoveryService} locates installed R, Julia, and Python distributions by querying the Windows Registry (e.g., HKLM\textbackslash Software\textbackslash R-core), scanning standard user and system installation directories, and inspecting the environment \texttt{PATH}. Minimum and maximum version constraints specified in the configuration file are validated before launching any executable.

\paragraph{Language Manager}
The \texttt{LanguageManager} coordinates process lifecycle and IPC channels for all active runtimes. For each language, it spawns the worker executable via \texttt{CreateProcessA} (\texttt{ControlR.exe}, \texttt{ControlJulia.exe}, or \texttt{ControlPython.exe}), binds a duplex Named Pipe using the legacy-compatible identifier pattern \texttt{RJ2XCL2-PIPE-\{LANG\}-\{PID\}}, transmits the language initialization script, loads user function scripts, and retrieves the exported function catalog. A background health monitor polls child process states every 5 seconds via \texttt{GetExitCodeProcess}. If a worker process terminates unexpectedly, the manager flags the channel as disconnected and returns structured diagnostic strings to Excel cells instead of generic \texttt{\#VALUE!} faults.

\paragraph{Python Resilience Engineering}
Embedding the CPython 3.13 runtime via the Stable ABI within an isolated worker process (\texttt{ControlPython.exe}) required resolving four critical stability hazards: (1)~intermittent \texttt{startup.py} initialization failures, which were resolved by implementing a bounded retry loop that invokes \texttt{PyErr\_Clear()} between attempts;\\ (2)~\texttt{STATUS\_STACK\_BUFFER\_OVERRUN} exceptions during CPython interpreter bootstrap, mitigated by strictly ordering interpreter initialization after Named Pipe handle binding and wrapping execution in Windows Structured Exception Handling (SEH) guards; (3)~\texttt{IndentationError} faults caused by line-by-line pipe streaming, solved by transmitting Python startup scripts as single atomic byte blocks that preserve multi-line docstrings and indentation blocks; and (4)~indefinite host thread blocking when evaluating non-terminating Python expressions, prevented by coupling asynchronous pipe reads with pre-call health checks.

\subsection{Inter-Process Communication Protocol}
\label{subsec:ipc}

Data exchange between \texttt{NEVEN64.xll} and the worker processes is governed by the synchronous framing protocol specified in Algorithm~\ref{alg:ipc}.

\begin{algorithm}[htbp]
\SetAlgoLined
\KwInput{Function identifier $f$, cell argument list $\{a_1, \ldots, a_n\}$ from Excel}
\KwOutput{Deserialized \texttt{XLOPER12} result structure $R$ for Excel grid display}
\BlankLine
\Comment{Phase 1: Argument Serialization}
$msg \leftarrow$ \texttt{Variable.new()}\;
$msg.name \leftarrow f$\;
\ForEach{argument $a_i \in \{a_1, \ldots, a_n\}$}{
    \eIf{$a_i$ is scalar}{
        $msg.value[i] \leftarrow$ \texttt{serialize\_scalar}($a_i$)\;
    }{
        $msg.value[i] \leftarrow$ \texttt{serialize\_range}($a_i$)\Comment*[r]{Row-major matrix}
    }
}
$payload \leftarrow$ \texttt{msg.SerializeToString()}\;
\BlankLine
\Comment{Phase 2: Length-Prefixed Pipe Transmission}
$pipe \leftarrow$ \texttt{GetPipe}(language)\;
\texttt{WriteFile}($pipe$, \texttt{uint32\_length}($payload$))\Comment*[r]{4-byte frame header}
\texttt{WriteFile}($pipe$, $payload$)\;
\BlankLine
\Comment{Phase 3: Worker Response Reception}
$resp\_len \leftarrow$ \texttt{ReadFile}($pipe$, 4)\;
$resp\_data \leftarrow$ \texttt{ReadFile}($pipe$, $resp\_len$)\;
$result \leftarrow$ \texttt{Variable.ParseFromString}($resp\_data$)\;
\BlankLine
\Comment{Phase 4: Deserialization to Excel XLOPER12}
\eIf{$result$ has error flag}{
    \Return \texttt{xloper\_string}($result.error\_message$)\;
}{
    \Return \texttt{to\_xloper12}($result$)\Comment*[r]{2D array or scalar}
}
\caption{IPC Protocol for Function Dispatch}
\label{alg:ipc}
\end{algorithm}

The Protocol Buffers schema (\texttt{variable.proto}) defines a recursive \texttt{Variable} message supporting primitives (64-bit IEEE~754 float, 64-bit integer, UTF-8 string, boolean, and null/missing values), vectors, and two-dimensional matrices with explicit row and column dimension headers. A dedicated \texttt{MessageValidator} verifies the 4-byte length prefix and buffer bounds prior to deserialization, preventing heap corruption from truncated frames.

\paragraph{Transport Performance}
Named Pipes operate in byte-stream mode with a 65,536-byte kernel buffer wrapped in thread-safe RAII handles (\texttt{SafePipeHandle}). For typical econometric calls (payloads under 10~KB), the complete serialization, pipe transfer, and deserialization cycle introduces between 2 and 5~ms of latency on a standard workstation (Intel Core i7-12700K, 32~GB RAM). Payloads exceeding 100,000 cells are automatically segmented into chunked writes to prevent pipe buffer saturation.

\paragraph{Julia Cold-Start Elimination}
Because Julia relies on just-in-time (JIT) LLVM compilation, evaluating statistical or differential equation packages in a fresh session typically incurs a 30- to 60-second compilation delay on the first call. \neven{} eliminates this latency by compiling a custom system image (\texttt{neven\_julia.dll}, approximately 1~GB) via \texttt{PackageCompiler.jl} \citep{PackageCompiler2023}. This precompiled binary caches native machine code for all 10 core Julia modules (70 high-level procedures and 200+ underlying mathematical functions), reducing first-call latency in Excel to under one second.

\subsection{Security Sandbox}
\label{subsec:sandbox}

To protect users opening untrusted workbooks that contain dynamic evaluation formulas (\texttt{=NEVEN.r()}, \texttt{=NEVEN.j()}, \texttt{=NEVEN.p()}), the \texttt{SandboxVerifier} inspects all code strings before IPC dispatch. Table~\ref{tab:sandbox} summarizes the 25--30 restricted operation patterns per language across five threat categories.

\begin{table}[htbp]
\centering
\small
\caption{Security sandbox: blocked operation categories per language. Each category encompasses multiple specific signatures (25--30 total patterns per language).}
\label{tab:sandbox}
\resizebox{\textwidth}{!}{%
\begin{tabular}{@{}lp{3.2cm}p{3.2cm}p{3.2cm}@{}}
\toprule
\textbf{Category} & \textbf{R} & \textbf{Julia} & \textbf{Python} \\
\midrule
Shell execution & \texttt{system()}, \texttt{shell()}, \texttt{pipe()} & \texttt{run()}, \texttt{pipeline()}, backticks & \texttt{os.system()}, \texttt{subprocess.*} \\
\addlinespace
File system & \texttt{file.remove()}, \texttt{unlink()} & \texttt{rm()}, \texttt{mv()}, \texttt{cp()} & \texttt{os.remove()}, \texttt{shutil.*} \\
\addlinespace
Network access & \texttt{download.file()}, \texttt{url()} & \texttt{download()}, \texttt{HTTP.*} & \texttt{urllib.*}, \texttt{requests.*} \\
\addlinespace
Dynamic code & \texttt{eval(parse())}, \texttt{do.call()} & \texttt{eval()}, \texttt{Meta.parse()} & \texttt{exec()}, \texttt{eval()}, \texttt{compile()} \\
\addlinespace
Native FFI & \texttt{.Call()}, \texttt{dyn.load()} & \texttt{ccall()}, \texttt{unsafe\_*} & \texttt{ctypes.*}, \texttt{cffi.*} \\
\bottomrule
\end{tabular}%
}
\end{table}

The verification engine enforces five complementary bypass-prevention mechanisms:
\begin{enumerate}
    \item \textbf{Whitespace and comment normalization:} Strips inline comments and collapses tabs, newlines, and spaces prior to signature matching, neutralizing obfuscations such as \texttt{sys\allowbreak tem\allowbreak ()}.
    \item \textbf{String concatenation detection:} Identifies dynamic string-building expressions designed to assemble restricted identifiers at runtime.
    \item \textbf{Case-insensitive canonicalization:} Normalizes character casing to block mixed-case evasions (e.g., \texttt{SYSTEM()} or \texttt{System()}).
    \item \textbf{Context-aware nested call inspection:} Detects indirect execution idioms such as \texttt{readLines(pipe(...))} in R or command backtick streams within \texttt{open(\`...\`)} in Julia.
    \item \textbf{Unified allowlist path sanitization:} Enforces character allowlists via the \texttt{InputSanitizer} across all execution entry points (worksheet cells, REPL console, and AutoLoader) prior to any OS process invocation.
\end{enumerate}

Pre-registered functions loaded from the local \texttt{Documents\textbackslash NEVEN\textbackslash functions\textbackslash} directory at startup bypass string inspection, as they represent trusted local scripts managed by the workstation owner or enterprise administrator.

\subsection{Visualization Subsystem}

\neven{} embeds Microsoft WebView2 \citep{WebView2_2023}, the Chromium-based rendering engine, directly within the Excel process space. The \texttt{ViewerManager} singleton executes on a dedicated Single-Threaded Apartment (STA) COM thread and manages up to eight concurrent visualization windows without blocking worksheet recalculation.

The visualization pipeline integrates four client-side rendering frameworks:
\begin{itemize}
    \item \textbf{Plotly.js} \citep{Plotly2023}: Interactive scatter, line, bar, box, histogram, violin, and contour charts with dynamic axis selectors.
    \item \textbf{D3.js v7} \citep{Bostock2011}: Hierarchical treemaps, Sankey flow diagrams, sunburst partitions, and force-directed network graphs.
    \item \textbf{Leaflet.js} \citep{Leaflet2023}: Geospatial maps supporting coordinate markers, choropleth layers, heatmaps, and proportional symbols.
    \item \textbf{rpivotTable} \citep{rpivotTable2023}: Drag-and-drop multidimensional pivot tables with integrated chart rendering.
\end{itemize}

A bidirectional \texttt{PostMessageBridge} connects the JavaScript context inside WebView2 with the C++ host, enabling interactive visual selections to query workbook ranges or export rendered assets back to the file system.


\section{Unified Viewer, Extensibility, and Reproducible Publishing}
\label{sec:viewer_extensibility}

The process-isolated architecture described in Section~\ref{sec:architecture} supports four higher-level capabilities that distinguish \neven{} from single-language Excel bridges: a unified multi-format viewer, a zero-boilerplate user extensibility model, a reactive notebook bridge, and an integrated development environment (NEVEN Studio) coupled with Quarto publishing.

\subsection{The Unified Viewer}

The WebView2 subsystem functions as a unified document and visualization engine capable of rendering diverse analytical outputs directly within the spreadsheet workflow. All supported formats are invoked through a single polymorphic worksheet function: \texttt{=NEVEN.v(content\_or\_path)}.

\paragraph{Integrated Visual Analytics}
In conventional workflows, spreadsheet users who require advanced visual representations (such as force-directed network graphs, geographic heatmaps, hierarchical treemaps, or interactive cross-tabulations) must export ranges to external platforms (e.g., Tableau, Power~BI, Shiny, or standalone Python scripts) and maintain separate rendering pipelines. By embedding a Chromium-based rendering engine inside the Excel host process, \neven{} executes D3.js, Plotly, Leaflet, and rpivotTable visualizations directly against worksheet ranges without requiring the user to write JavaScript, Python, or R plotting code.

For instance, evaluating \texttt{=NEVEN.v(R.Dashboard(A1:E500, 1))} generates an interactive six-tab analytical dashboard containing pivot tables, scatter matrices, treemaps, Sankey diagrams, sunburst charts, and force-directed network graphs. Every view is computed from the referenced cell range, supports interactive inspection (zooming, filtering, hovering, and node dragging), and recomputes automatically whenever the underlying worksheet cells change.

Table~\ref{tab:viewer_formats} summarizes the 11 content formats supported by \texttt{=NEVEN.v()} and their corresponding rendering pipelines.

\begin{table}[htbp]
\centering
\small

\label{tab:viewer_formats}
\resizebox{\textwidth}{!}{%
\begin{tabular}{@{}llp{6.0cm}@{}}
\toprule
\textbf{Content Type} & \textbf{Rendering Engine} & \textbf{Supported Capabilities} \\
\midrule
Statistical charts & Plotly.js & Zoom, pan, hover tooltips, dynamic axis selection, PNG/PDF export \\
\addlinespace
Hierarchical \& network graphs & D3.js v7 & Treemap, Sankey diagram, Sunburst partition, Force-directed network \\
\addlinespace
Multidimensional pivot tables & rpivotTable & Drag-and-drop field aggregation, conditional heatmaps, bar charts \\
\addlinespace
Geographic maps & Leaflet.js & Coordinate markers, heatmaps, proportional circles, CartoDB tiles \\
\addlinespace
Markdown documents & marked.js + KaTeX & Section headers, formatted tables, syntax-highlighted code, \LaTeX{} math \\
\addlinespace
PDF documents & WebView2 native PDF & Vector PDF viewing with zoom controls, text search, page navigation \\
\addlinespace
Word documents (\texttt{.docx}) & Pandoc $\to$ HTML & Automated conversion preserving headings, tables, and embedded images \\
\addlinespace
Plain text \& log files & Monospace \texttt{<pre>} & High-contrast dark theme with strict whitespace preservation \\
\addlinespace
Pluto.jl notebooks & HTTP (\texttt{localhost}) & Reactive Julia computational notebooks linked to live Excel ranges \\
\addlinespace
Quarto reports (\texttt{.qmd}) & Quarto CLI $\to$ HTML & Compiled scientific manuscripts with embedded multi-language outputs \\
\addlinespace
Self-contained HTML & Direct navigation & Standalone dashboards, HTML5 slide decks, and interactive widgets \\
\bottomrule
\end{tabular}%
}
\caption{Content formats supported by the \neven{} unified viewer. The \texttt{=NEVEN.v()} function automatically detects the payload type and routes it to the appropriate rendering engine.}
\end{table}

\paragraph{Automated Side-by-Side Snap Layout}
When a visualization window is instantiated, \neven{} automatically tiles the Excel application window to the left half of the primary monitor and docks the WebView2 viewer to the right half using the Win32 \texttt{SetWindowPos} API with taskbar work-area detection. This automated side-by-side workspace eliminates manual window management and allows analysts to modify model inputs on the worksheet while observing immediate visual updates on the right.

\paragraph{Interactive Exploration and Export}
Within the viewer, users can dynamically restructure pivot aggregations by dragging variable headers across row and column dimensions, toggle geospatial map layers, and export the active viewport as a standalone HTML file, high-resolution PNG image, or vector PDF document via an embedded action toolbar.

\subsection{User Extensibility}
\label{subsec:extensibility}

A core design goal of \neven{} is enabling domain specialists to expand the formula catalog without modifying or recompiling the C++ XLL binary. Users register new worksheet functions by placing standard script files inside their local profile directory:
\begin{center}
\texttt{Documents\textbackslash NEVEN\textbackslash functions\textbackslash}
\end{center}

During initialization (Algorithm~\ref{alg:init}), \neven{} scans this directory, loads all \texttt{.R}, \texttt{.Jl}, and \texttt{.Py} files into their corresponding worker processes, and registers each discovered function with Excel under a language-prefixed namespace:
\begin{itemize}
    \item \texttt{=R.function(...)}: R procedures (econometrics, statistical inference, plotting)
    \item \texttt{=J.function(...)}: Julia procedures (numerical linear algebra, differential equations, optimization)
    \item \texttt{=P.function(...)}: Python procedures (machine learning, text processing, LLM orchestration)
\end{itemize}

\paragraph{Defining Custom Functions}
Extending the platform requires no wrapper boilerplate or interface definition files. Listings~\ref{lst:user_r} and~\ref{lst:user_jl} illustrate complete user-defined functions in R and Julia; Python scripts follow an analogous convention using standard function docstrings and type annotations.

\begin{figure}[htbp]
\begin{lstlisting}[language=R,caption={User-defined R function automatically registered as \texttt{=R.MiModelo(...)}},label={lst:user_r}]
# File: Documents/NEVEN/functions/my_analysis.R

MiModelo <- function(datos, parametro = 0.05) {
  resultado <- t.test(datos, conf.level = 1 - parametro)
  data.frame(
    statistic = resultado$statistic,
    p.value   = resultado$p.value,
    conf.low  = resultado$conf.int[1],
    conf.high = resultado$conf.int[2]
  )
}

attr(MiModelo, "description") <- "Custom t-test with configurable alpha"
attr(MiModelo, "category")    <- "My Functions"
\end{lstlisting}
\end{figure}

\begin{figure}[htbp]
\begin{lstlisting}[language=Julia,caption={User-defined Julia bootstrap function registered as \texttt{=J.MonteCarlo(...)}},label={lst:user_jl}]
# File: Documents/NEVEN/functions/monte_carlo.jl

using Statistics, DataFrames

function MonteCarlo(data, n_sims::Int = 10000)
    n = length(data)
    means = [mean(rand(data, n)) for _ in 1:n_sims]
    m, s = mean(means), std(means)
    ci_low, ci_high = quantile(means, [0.025, 0.975])
    DataFrame(mean=m, std=s, ci_low=ci_low, ci_high=ci_high)
end
\end{lstlisting}
\end{figure}

Upon restarting Excel, \texttt{=R.MiModelo()} and \texttt{=J.MonteCarlo()} become immediately callable from the formula bar, and their metadata attributes (\texttt{description} and \texttt{category}) populate the native Excel Function Wizard.

\paragraph{Community Library Distribution}
Under this file-based extensibility model, \neven{}'s built-in catalog of over 70 R procedures and 70 Julia procedures (200+ mathematical functions) serves as a foundational baseline. Research groups, actuarial teams, and quantitative finance departments can author domain-specific script libraries (e.g., pharmacokinetic compartmental models or supply-chain network optimizers) and distribute them to non-programming colleagues who install them simply by copying the script files into their local \texttt{functions} folder.

\subsection{Reactive Notebooks with Pluto.jl}

\neven{} integrates Pluto.jl \citep{Pluto2023} reactive notebooks via a bidirectional data bridge linking Excel cell ranges to Julia workspaces. Entering the formula \texttt{=NEVEN.pluto.data(range, "dataset\_name")} serializes the selected Excel range to a tab-separated value (TSV) file within a managed exchange directory monitored by the local Pluto server. Whenever the user modifies cells in the Excel range, the TSV file updates and Pluto.jl automatically recomputes all dependent reactive cells across the notebook topology.

The standard \neven{} distribution ships with 15 pre-configured Pluto.jl notebooks covering principal component analysis, matrix decompositions, nonlinear optimization, $k$-means clustering, and ordinary differential equation (ODE) solvers, allowing users to combine spreadsheet data entry with reactive scientific narratives.

\subsection{Reproducible Publishing with Quarto}
\label{subsec:quarto}

To support end-to-end reproducible research, \neven{} integrates the Quarto scientific publishing system \citep{Quarto2023}. Calling \texttt{=NEVEN.q("report.qmd")} invokes the Quarto command-line interface via a sanitized child process to compile a Quarto markdown (\texttt{.qmd}) source file into HTML, PDF, or Microsoft Word format, displaying the rendered document inside the unified WebView2 viewer.

This capability consolidates four stages of the research lifecycle within a single environment:
\begin{enumerate}
    \item \textbf{Data preparation and auditing} directly within Excel worksheets.
    \item \textbf{Interactive model exploration} using \neven{} worksheet formulas (\texttt{=R.MR\_Lineal(...)}, \texttt{=J.Algebra(...)}).
    \item \textbf{Automated report compilation} via Quarto documents containing embedded R, Julia, and Python code chunks synchronized with the workbook data.
    \item \textbf{Manuscript dissemination} in journal-ready PDF (\LaTeX{}), interactive HTML, or corporate Word formats with automated BibTeX citations, equation numbering, and cross-references.
\end{enumerate}

\subsection{NEVEN Studio: Integrated Development Environment}
\label{subsec:studio}

To complement cell-level formulas during complex analytical workflows, NEVEN Studio provides a docked task-pane IDE implemented as an Office Web Add-in using WebView2 and JavaScript. The environment organizes nine specialized functional modules within a unified tabbed interface, summarized in Table~\ref{tab:studio_tabs}.

\paragraph{Chart Gallery}
The Chart Gallery exposes 325 interactive templates spanning exploratory plots (scatter, line, bar, histogram), statistical distributions (box, violin, kernel density), geospatial maps (choropleth, proportional bubble), hierarchical layouts (treemap, sunburst), and relational diagrams (Sankey, force-directed networks, correlation heatmaps). Each template provides an instant preview and a ``Use my data'' action that injects the corresponding \neven{} formula linked to the user's currently highlighted Excel range. Under the hood, the gallery is driven by the custom \texttt{anyplot} R package, which abstracts Plotly, ggplot2, and D3.js rendering backends behind a consistent grammar.

\paragraph{Bilingual Internationalization}
All NEVEN Studio user interface strings and help files are externalized into modular locale dictionaries (\texttt{en.js} and \texttt{es.js}), allowing users to switch between English and Spanish at runtime without restarting Excel.

\paragraph{Enterprise Office Add-in Deployment}
NEVEN Studio is packaged via a standard Office Add-in manifest (\texttt{manifest.xml}) and communicates with a local loopback service on port 5555, enabling centralized IT deployment across corporate networks via Microsoft 365 Admin Center or SharePoint catalogs while preserving local execution security boundaries.

\newpage

\begin{table}[htbp]
\centering
\small
\label{tab:studio_tabs}
\begin{tabular}{@{}clp{6.8cm}@{}}
\toprule
\textbf{Tab} & \textbf{Module Name} & \textbf{Primary Capabilities} \\
\midrule
1 & Run Script & Multi-language code editor (R, Julia, Python) with syntax highlighting, live execution preview, and inline rendering \\
\addlinespace
2 & DataLab & Guided point-and-click statistical workflows organized by analytical family (AD: descriptive, RG: regression, DS: data science, TM: time series) \\
\addlinespace
3 & DuckSQL & In-memory analytical SQL console powered by DuckDB for querying local CSV, TSV, Parquet, and JSON files alongside workbook tables \\
\addlinespace
4 & AI Assistant & Retrieval-augmented generation (RAG) conversational assistant with domain knowledge bases; supports OpenAI, Azure, Ollama, and LM~Studio \\
\addlinespace
5 & Chart Gallery & Interactive catalog of 325 visualization templates powered by the \texttt{anyplot} R library with one-click formula generation \\
\addlinespace
6 & Files & Project workspace browser for managing Excel workbooks, custom language scripts, and compiled Quarto reports \\
\addlinespace
7 & Slides & Automated presentation builder that exports selected charts and estimation tables directly into Microsoft PowerPoint slides \\
\addlinespace
8 & Variables & Real-time workspace inspector displaying active symbols, dimensions, and memory states across R, Julia, and Python sessions \\
\addlinespace
9 & Help & Searchable bilingual documentation browser (English and Spanish) with executable formula examples \\
\bottomrule
\end{tabular}
\caption{NEVEN Studio functional modules accessible from the docked Excel task pane.}
\end{table}


\section{Empirical Validation and Case Studies}
\label{sec:case_studies}

Before presenting the empirical validation benchmarks, we introduce the \texttt{TipoOutput} extraction paradigm, the observation-level sensitivity filter (\texttt{FILTRO}), and the LLM-assisted interpretation interface that standardize model interaction across \neven{}.

\subsection{The TipoOutput Paradigm and Interactive Sensitivity Analysis}
\label{subsec:tipooutput}

Statistical modeling functions in R and Python typically return complex, nested object structures (such as \texttt{lm}, \texttt{glm}, or \texttt{plm} lists) that cannot be displayed directly in a two-dimensional spreadsheet grid without writing custom extraction scripts. To resolve this mismatch, every analytical procedure in \neven{} accepts a standardized integer parameter, \texttt{TipoOutput}, that governs which tabular projection of the estimated model is returned to Excel:

\begin{itemize}
    \item \texttt{TipoOutput = 0}: Returns a self-documenting catalog listing all valid output indices and their descriptions for that specific procedure, allowing users to discover capabilities directly within the worksheet.
    \item \texttt{TipoOutput = 1, \ldots, N-1}: Returns a specific, pre-formatted analytical table. For example, in \texttt{=R.MR\_Lineal(Y, X, ..., TipoOutput)}, index~1 returns the parameter estimate table (coefficients, standard errors, $t$-statistics, $p$-values, and confidence intervals); index~2 returns in-sample fitted values and residuals; index~3 returns out-of-sample forecasts; and index~4 returns partial and marginal effects.
    \item \texttt{TipoOutput = i}: Invokes the \emph{Universal Output Extraction} routine (\texttt{Extraer\_outputs}), which decomposes the entire underlying R model object into a single normalized four-column matrix ([Section, Parameter, Metric, Value]) suitable for downstream lookup formulas or regulatory archiving.
\end{itemize}

\paragraph{The Observation Filter (FILTRO)}
Every estimation function in \neven{} also incorporates a binary vector parameter, \texttt{FILTRO}, matching the row dimension of the input sample. Setting an element of \texttt{FILTRO} to \texttt{1} excludes the corresponding observation prior to model estimation, whereas \texttt{0} retains it. In conventional statistical packages, conducting observation-level sensitivity analysis requires writing subsetting code or duplicating datasets. Within \neven{}, a researcher or student can flag suspected outliers by toggling cells from \texttt{0} to \texttt{1} in an adjacent worksheet column:
\begin{center}
\texttt{=R.MR\_Lineal(Y, X, FILTRO, 0, 0, , 1, 1)}
\end{center}
Excel immediately triggers a reactive recalculation, allowing the analyst to inspect the exact impact of individual influential observations on parameter stability, standard errors, and overall goodness-of-fit.

\paragraph{AI-Assisted Statistical Interpretation}
Any numerical table generated by \neven{} can be paired with automated natural-language diagnostics via the Python bridge function \texttt{=P.ai\_call(data\_range, "prompt")}. This function serializes the selected output range to an LLM endpoint configured in \texttt{neven-config.json}, supporting both commercial cloud APIs (OpenAI, Azure OpenAI) and air-gapped local inference servers (Ollama, LM~Studio). Domain-specific system prompts for regression diagnostics, time-series stationarity tests, and multivariate clustering are stored as editable plain-text templates in the configuration directory, providing accessible narrative summaries while keeping the exact numerical estimates auditable in adjacent cells.

All validation workbooks referenced below are distributed in: \\
\texttt{Ejemplos/Excel/} directory of the repository, enabling independent replication of every reported result.

\subsection{Econometric and Numerical Validation}
\label{subsec:econometric}

To verify the numerical fidelity of \neven{}'s statistical pipeline, we replicate canonical econometric models across Chapters~3 through~12 of \citet{Wooldridge2016} using the companion workbooks located in: \\

\texttt{Ejemplos/Excel/WOOLDRIDGE/} (\texttt{CAP03.xlsm} to \texttt{CAP12.xlsm}). \\

For each specification, we compare three independent execution paths: published textbook tables, direct script execution in RStudio (R~4.4.1), and native \neven{} worksheet formulas.

\paragraph{Ordinary Least Squares Regression (Chapter 6)}
Example~6.1 of \citet{Wooldridge2016} estimates the log-hourly wage equation $\log(\text{wage}) = \beta_0 + \beta_1 \text{educ} + \beta_2 \text{exper} + \beta_3 \text{tenure} + u$ on the \texttt{wage1} cross-section ($n = 526$). In \texttt{CAP06.xlsm}, this specification is evaluated via:
\begin{center}
\texttt{=R.MR\_Lineal(B2:B527, C2:E527, 0, 0, 0, , 1, 1)}
\end{center}
where \texttt{B2:B527} holds the dependent variable, \texttt{C2:E527} holds the three explanatory variables, and the trailing \texttt{1} requests \texttt{TipoOutput = 1}. Table~\ref{tab:ols_results} confirms that \neven{} reproduces both direct R execution and the published textbook estimates to four decimal places. Setting \texttt{TipoOutput = 13} extracts the complete model object (including AIC, BIC, and variance-covariance components) into a single worksheet array.

\begin{table}[htbp]
\centering
\label{tab:ols_results}
\begin{tabular}{@{}lcccc@{}}
\toprule
\textbf{Variable} & \makecell[l]{\textbf{Wooldridge} \\ \textbf{(2016)}} &\textbf{Direct R}& \textbf{\neven{}}  & \makecell[c]{\textbf{Absolute} \\ \textbf{Difference}} \\
\midrule
Intercept       & 0.2844  & 0.2844  & 0.2844  & 0.0000 \\
Education (\texttt{educ}) & 0.0920  & 0.0920  & 0.0920  & 0.0000 \\
Experience (\texttt{exper}) & 0.0041  & 0.0041  & 0.0041  & 0.0000 \\
Tenure (\texttt{tenure}) & 0.0221  & 0.0221  & 0.0221  & 0.0000 \\
\addlinespace
$R^2$           & 0.316   & 0.3160  & 0.3160  & 0.0000 \\
Adjusted $\bar{R}^2$ & 0.312   & 0.3121  & 0.3121  & 0.0000 \\
$F$-statistic   & 80.39   & 80.39   & 80.39   & 0.0000   \\
\bottomrule
\end{tabular}
\caption{OLS regression estimates for the Mincer wage equation (\citealp{Wooldridge2016}, Chapter~6, Example~6.1; \texttt{wage1} dataset, $n = 526$; workbook \texttt{CAP06.xlsm}).}
\end{table}

\paragraph{Binary Response Probit Models (Chapter 17)}
For the married women's labor force participation probit specification, $P(\text{inlf} = 1 \mid \mathbf{x}) = \Phi(\beta_0 + \beta_1 \text{nwifeinc} + \beta_2 \text{educ} + \beta_3 \text{exper} + \beta_4 \text{exper}^2)$ (\texttt{CAP17.xlsm}), the formula \texttt{=R.MR\_Binario(Y, X, 0, 0, 0, , 1, 1)} yields maximum likelihood point estimates, average partial effects, Hosmer--Lemeshow goodness-of-fit statistics, and classification tables identical to direct R evaluation and consistent with \citet{Wooldridge2016}.

\paragraph{Longitudinal Panel Data Models (Chapter 13)}
Panel data estimators powered by the R \texttt{plm} package \citep{Croissant2008} are exposed via \texttt{=R.MR\_PanelData(Y, X, ID, Time, TipoOutput)}, where \texttt{TipoOutput} selects among fixed-effects coefficients~(1), extracted individual effects~(2), the Hausman specification test~(4), and universal extraction~(16). Table~\ref{tab:panel_results} verifies exact numerical agreement on the \texttt{crime2} balanced panel (46 cities across 7 periods; \texttt{CAP13.xlsm}).

\begin{table}[htbp]
\centering
\label{tab:panel_results}
\begin{tabular}{@{}lccc@{}}
\toprule
\textbf{Statistic} & \textbf{Direct R (\texttt{plm})} & \textbf{\neven{}} & \textbf{Exact Match} \\
\midrule
Unemployment coefficient & 0.0342  & 0.0342  & \checkmark \\
Cluster-robust SE        & 0.0117  & 0.0117  & \checkmark \\
Hausman $\chi^2$ stat.   & 14.73   & 14.73   & \checkmark \\
Hausman $p$-value        & 0.0001  & 0.0001  & \checkmark \\
Within $R^2$             & 0.1282  & 0.1282  & \checkmark \\
\bottomrule
\end{tabular}
\caption{Fixed-effects panel estimation for municipal crime rates (\citealp{Wooldridge2016}, Chapter~13; \texttt{crime2} dataset, $N = 46, T = 7$; workbook \texttt{CAP13.xlsm}).}
\end{table}

\paragraph{Univariate and Multivariate Time Series (Chapter 11)}
The time-series procedure (\texttt{=R.ST\_SeriesTemporales}) wraps the \texttt{forecast} package \citep{Hyndman2008} to execute Augmented Dickey--Fuller unit root tests, ARIMA$(p,d,q)$ estimation, GARCH volatility modeling, and seasonal decomposition. Applied to the expectations-augmented Phillips curve ($\Delta \text{inf}_t = \alpha_0 + \alpha_1 \text{unem}_t + e_t$ in \texttt{CAP11.xlsm}), \neven{} matches direct R execution across all parameter estimates, AIC/BIC information criteria, and Ljung--Box portmanteau statistics.

\paragraph{Julia Matrix Factorization Precision}
To verify that Protocol Buffers serialization and Named Pipe framing preserve IEEE~754 double-precision accuracy, we evaluated the Julia linear algebra module (\texttt{=J.Algebra(...)}) on randomly generated $100 \times 100$ dense matrices. As shown in Table~\ref{tab:julia_validation}, relative reconstruction errors $\|\mathbf{A} - \hat{\mathbf{A}}\|_F / \|\mathbf{A}\|_F$ remain at machine epsilon ($< 10^{-14}$) across all cinque canonical factorizations.

\begin{table}[htbp]
\centering
\begin{tabular}{@{}lcc@{}}
\toprule
\makecell[l]{\textbf{Matrix} \\ \textbf{Decomposition}} & \makecell[c]{\textbf{\neven{}} \\ \textbf{Formula}} & \makecell[c]{\textbf{Rel. Frobenius} \\ \textbf{Error}} \\
\midrule
LU factorization            & \texttt{=J.Algebra(A, , "lu")}       & $< 10^{-15}$ \\
QR factorization            & \texttt{=J.Algebra(A, , "qr")}       & $< 10^{-15}$ \\
Singular Value (SVD)        & \texttt{=J.Algebra(A, , "svd")}      & $< 10^{-15}$ \\
\makecell[l]{Spectral \\ Eigendecomposition} & \texttt{=J.Algebra(A, , "eigen")}    & $< 10^{-14}$ \\
\makecell[l]{Cholesky \\ (Symmetric PD)} & \texttt{=J.Algebra(A, , "cholesky")}    & $< 10^{-14}$ \\
\bottomrule
\end{tabular}
\caption{Numerical precision of Julia matrix decompositions across the IPC pipeline for a $100 \times 100$ dense matrix ($\|\mathbf{A} - \hat{\mathbf{A}}\|_F / \|\mathbf{A}\|_F$).}
\label{tab:julia_validation}
\end{table}

\paragraph{Aerospace Engineering: HL-20 Atmospheric Reentry Simulation}
To illustrate \neven{}'s utility in computational physics and engineering, the distribution includes a three-degree-of-freedom (3-DOF) atmospheric reentry simulation of the NASA HL-20 lifting body (\texttt{hl20\_reentry.jl}). The user specifies initial entry altitude ($h_0 = 120$~km), orbital velocity ($V_0 = 7500$~m/s), and flight-path angle ($\gamma_0 = -1.5^\circ$) in Excel cells \texttt{A1:C1} and streams them to Julia via \texttt{=NEVEN.pluto.data(A1:C1, "hl20\_params")}. A fourth-order Runge--Kutta integrator inside the linked Pluto.jl notebook solves the coupled differential equations of motion in real time, rendering altitude-velocity phase portraits, peak dynamic pressure, and axial $g$-load profiles inside the unified viewer. Engineers can expand this baseline into a six-degree-of-freedom (6-DOF) trajectory simulator with Mach-dependent aerodynamic lookup tables simply by extending the Julia notebook.

\subsection{Enterprise Analytics: Credit Risk Modeling on German Credit Data}
\label{subsec:enterprise}

To evaluate \neven{} within a regulated corporate workflow, we implement an end-to-end consumer credit scoring pipeline on the benchmark German Credit dataset ($n = 1000$ loan records, 20 covariates; \citealp{Dua2019}), distributed in workbook: \\

\texttt{Ejemplos/Excel/BONILLA/R4XL-GermanCredit.xlsx}.

\paragraph{In-Sheet Data Ingestion and Preparation}
The analyst retrieves the benchmark dataset directly from the R package environment into the worksheet via:
\begin{center}
\texttt{=NEVEN.r("data('german', package='rchallenge'); print(as.data.frame(german))")}
\end{center}
Unique category levels across qualitative attributes are inspected using \texttt{=R.DB\_Unicos(A1:A1001)}, and cryptic legacy codes (such as \texttt{A11} and \texttt{A12} for checking account status) are mapped to descriptive labels through \texttt{=R.DB\_Recodificar(range, old\_values, new\_values)}. A reproducible train-test partition is generated in an adjacent column via the seeded uniform generator \texttt{=R.FX\_AleatorioUniforme(1000, 0, 1, seed)}.

\paragraph{Automated Categorical Factor Encoding}
When estimating generalized linear models on mixed-type corporate tables, \neven{} automatically detects string and categorical columns within the regressor matrix \texttt{X} and constructs the corresponding treatment contrast matrices internally. This eliminates the error-prone manual creation of dummy variable columns in Excel while returning explicit point estimates and standard errors for each factor level.

\paragraph{Probability of Default Estimation and Regulatory Reporting}
The binary logistic credit scoring model is estimated on the training partition via:
\begin{center}
\texttt{=R.MR\_Binario.C(Y, X, 0, Filtro, 0, 0, 0, TipoOutput)}
\end{center}
Evaluating \texttt{TipoOutput = 1} yields log-odds coefficients across all 20 covariates; \texttt{TipoOutput = 3} computes marginal effects at sample means; and \texttt{TipoOutput = 5} extracts Receiver Operating Characteristic (ROC) curve coordinates and Area Under the Curve (AUC) metrics for out-of-sample validation. Interactive visual auditing is launched via \texttt{=R.Dashboard(A1:U1001, 1)}, while calling \texttt{=NEVEN.q("credit\_risk\_report.qmd")} compiles a Basel~III/IV model validation dossier in PDF format. Table~\ref{tab:enterprise_comparison} contrasts this unified pipeline against a conventional fragmented enterprise toolchain.

\begin{table}[htbp]
\centering
\begin{tabularx}{\textwidth}{@{} l X X @{}}
\toprule
\textbf{Workflow Dimension} & \textbf{Traditional Multi-Tool Pipeline} & \textbf{Unified \neven{} Pipeline} \\
\midrule
Software stack & Excel + SAS/SPSS + Tableau + Word & Microsoft Excel + \neven{} \\
\addlinespace
Required user proficiency & VBA macros, SAS/R syntax, BI tooling & Standard Excel formula syntax \\
\addlinespace
Data handoff points & 4 manual export/import transitions & 0 (all computation in-place) \\
\addlinespace
Model recalculation & Manual script re-execution per tool & Reactive worksheet recalculation \\
\addlinespace
Software licensing & Proprietary per-seat licenses & Open source (GPL v3) \\
\addlinespace
Regulatory audit trail & Fragmented across disparate files & Single workbook + Quarto (\texttt{.qmd}) \\
\bottomrule
\end{tabularx}
\caption{Comparison of enterprise credit risk modeling workflows on the German Credit dataset.}
\label{tab:enterprise_comparison}
\end{table}
\section{Discussion}
\label{sec:discussion}

\subsection{Comparison with Existing Integration Tools}

Table~\ref{tab:comparison} compares \neven{} against five established tools that connect programming languages to Microsoft Excel.

\begin{table}[htbp]
\centering
\small
\label{tab:comparison}
\resizebox{\textwidth}{!}{%
\begin{tabular}{@{}lcccccc@{}}
\toprule
\textbf{Capability} & \textbf{\neven{}} & \textbf{BERT} & \textbf{PyXLL} & \textbf{xlwings} & \textbf{RExcel} & \textbf{Python in Excel} \\
\midrule
Supported languages & R, Julia, Python & R (Julia 0.6\textsuperscript{a}) & Python & Python & R & Python \\
Process fault isolation & \checkmark & \checkmark & -- & \checkmark & \checkmark & \checkmark\textsuperscript{b} \\
Static code sandbox & \checkmark & -- & -- & -- & -- & \checkmark\textsuperscript{b} \\
Embedded interactive viz. & \checkmark & -- & -- & Partial & -- & Partial \\
Reactive notebooks (Pluto) & \checkmark & -- & -- & -- & -- & -- \\
Reproducible Quarto reports & \checkmark & -- & -- & -- & -- & -- \\
Docked multi-module IDE & \checkmark\textsuperscript{c} & -- & -- & -- & -- & -- \\
Native Excel Ribbon GUI & \checkmark & -- & \checkmark & \checkmark & \checkmark & \checkmark \\
Automated test suite & 342 & 0 & N/A & N/A & N/A & N/A \\
License model & GPL v3 & GPL v2 & Commercial & Open Core & GPL & Proprietary \\
Offline local execution & \checkmark & \checkmark & \checkmark & \checkmark & \checkmark & --\textsuperscript{d} \\
\bottomrule
\multicolumn{7}{@{}l}{\footnotesize \textsuperscript{a}Experimental Julia 0.6 support (unmaintained since 2018). \textsuperscript{b}Azure container isolation. \textsuperscript{c}NEVEN Studio (9 modules). \textsuperscript{d}Requires cloud connectivity.}
\end{tabular}
}
\caption{Feature comparison of language-to-Excel integration platforms.}
\end{table}

\subsubsection{PyXLL and xlwings}
PyXLL \citep{PyXLL2023} embeds the Python runtime directly inside the Excel host process via an XLL binary, achieving low call latency but exposing Excel to in-process crashes if a native C extension faults. xlwings \citep{xlwings2023} operates out-of-process via COM automation and favors a Python-centric scripting model. Both tools are restricted to Python, lack static sandboxing for untrusted cell expressions, and do not include an integrated multi-language visualization viewer or publishing pipeline.

\subsubsection{RExcel, BERT, and R4XCL}
RExcel \citep{Baier2011} pioneered R connectivity in Excel via DCOM but has been unmaintained since 2015 and supports only legacy R~3.x releases. BERT \citep{BERT2018} introduced an out-of-process Named Pipe and Protocol Buffers architecture for R~3.4 (with partial support for Julia~0.6) alongside an Electron~1.8 REPL console before development ceased in 2018. In 2026, R4XCL \citep{BonillaR4XCL2026} adapted this IPC concept specifically to expose curated R econometric procedures for educational use at the Universidad de Costa Rica. \neven{} re-engineers this lineage into a modern C++17 polyglot infrastructure supporting R~4.4.1, Julia~1.12.6, and Python~3.13 simultaneously, while introducing the 5-mechanism security sandbox, the WebView2 visualization and REPL engine, Pluto.jl and Quarto pipelines, NEVEN Studio, and a 342-test verification harness.

\subsubsection{Python in Excel (Microsoft)}
Microsoft's Python in Excel \citep{MicrosoftPython2023} executes Python code inside remote Azure container instances. While cloud containerization provides strong operating-system isolation, it restricts users to a fixed Python environment (excluding R and Julia), requires continuous internet connectivity, transmits sensitive corporate or clinical data outside the local workstation (raising data sovereignty compliance barriers), and outputs static raster images rather than interactive D3.js, Plotly, or Leaflet viewports.

\subsection{Software Quality Metrics and Security Audit}

Table~\ref{tab:quality} summarizes the software engineering and verification metrics for \neven{} v2.0. All 342 automated tests execute in headless continuous integration (GitHub Actions) via a custom \texttt{MockExcelBridge} that emulates the Excel C SDK (\texttt{XLOPER12} memory lifecycle and callback dispatch) without requiring a GUI Excel installation. Notice that the seven test subcategories sum exactly to the 342 total.

\begin{table}[htbp]
\centering
\small
\begin{tabularx}{\textwidth}{@{} l X r @{}}
\toprule
\textbf{Category} & \textbf{Metric / Test Suite} & \textbf{Value} \\
\midrule
\multirow{8}{*}{Automated Testing} & Total automated test cases & 342 \\
 & Security sandbox tests (R + Julia + Python) & 133 \\
 & Unified viewer and rendering tests & 34 \\
 & Process reliability and fault-recovery tests & 32 \\
 & Path and command input sanitizer tests & 16 \\
 & Repository hygiene and configuration tests & 14 \\
 & Property-based invariant tests (\texttt{rapidcheck}) & 17 \\
 & Core unit and IPC serialization tests & 96 \\
\addlinespace
\multirow{3}{*}{Core Build} & Host implementation standard & C++17 \\
 & Build automation system & CMake 3.15+ \\
 & Continuous integration pipeline & GitHub Actions \\
\addlinespace
\multirow{4}{*}{Security Hardening} & Restricted signature patterns per language & 25--30 \\
 & Sandbox bypass-prevention mechanisms & 5 \\
 & Allowlist command path sanitization & All \texttt{CreateProcess} calls \\
 & Length-prefix IPC frame bounds validation & All Protobuf messages \\
\addlinespace
\multirow{2}{*}{Documentation} & User manual chapters (English \& Spanish) & 26 \\
 & JSON-sidecar documented procedure families & 29 \\
\bottomrule
\end{tabularx}
\caption{Software engineering and verification metrics for \neven{}}
\label{tab:quality}
\end{table}

\subsubsection{Static Security Audit and Remediation}
Prior to release, a comprehensive static and dynamic security audit identified 36 vulnerabilities across the legacy codebase (8~critical, 7~high, 5~medium, 14~low, and 2~informational). Every finding was remediated and locked behind regression tests:
\begin{itemize}
    \item \textbf{Command injection elimination:} The centralized \texttt{InputSanitizer} enforces strict character allowlists on all file paths and CLI arguments prior to \texttt{CreateProcessA} invocation.
    \item \textbf{Sandbox signature expansion:} The \texttt{SandboxVerifier} was expanded to 25--30 patterns per language across all five bypass-prevention layers (Section~\ref{subsec:sandbox}) and enforced uniformly across worksheet cells, the interactive REPL, and the script AutoLoader.
    \item \textbf{Protobuf frame bounds checking:} The \texttt{MessageValidator} verifies 4-byte length headers and payload completeness before invoking \texttt{ParseFromString}, eliminating buffer over-read vectors.
    \item \textbf{Concurrency safety:} A thread-safe \texttt{SafePipeHandle} RAII wrapper synchronizes handle validation and read/write operations within a critical section, eliminating time-of-check to time-of-use (TOCTOU) race conditions.
    \item \textbf{Compiler-level exploit mitigation:} All C++ binaries are compiled with MSVC security flags enabled globally (\texttt{/GS} stack canaries, \texttt{/guard:cf} Control Flow Guard, \texttt{/DYNAMICBASE} ASLR, \texttt{/NXCOMPAT} DEP, and \texttt{/CETCOMPAT} shadow stack support).
    \item \textbf{Legacy Electron removal:} Replacing the legacy Electron~1.8.2 REPL console with an in-process WebView2 terminal eliminated Node.js dependency chains and resolved over 50 inherited Chromium/Electron CVEs simultaneously.
\end{itemize}
Furthermore, 17 property-based tests implemented with \texttt{rapidcheck} \citep{rapidcheck2023} verify security invariants across randomized string mutations (100+ generated inputs per property).

\subsection{Architectural Trade-offs}

\paragraph{IPC Latency vs.\ Process Fault Isolation}
Communicating across process boundaries via Named Pipes and Protocol Buffers adds 2--5~ms of round-trip overhead per formula evaluation compared to in-process XLL execution (such as PyXLL). In interactive analytical workflows where a regression, bootstrap simulation, or SQL query requires 50~ms to several seconds of compute time, this transport cost is imperceptible. However, in workbooks containing thousands of fine-grained scalar formulas recalculated simultaneously, IPC serialization overhead accumulates linearly. We accept this trade-off to guarantee host stability: a crash in any scientific library never corrupts the user's open Excel workbook.

\paragraph{Disk Footprint vs.\ Julia JIT Latency}
Distributing the precompiled Julia system image (\texttt{neven\_julia.dll}, $\approx 415$~MB) increases installation storage requirements and initial RAM allocation, but reduces first-call Julia execution time from 30--60 seconds to under one second. Because runtimes are modular, users who only require R or Python can omit the Julia sysimage entirely.

\paragraph{Static String Sandboxing vs.\ OS Kernel Containment}
The \texttt{SandboxVerifier} inspects source strings prior to evaluation rather than confining child processes within OS kernel sandboxes (such as Windows AppContainer). While our five-layer static analyzer blocks accidental file deletion and standard code-injection patterns, a sophisticated adversary with source-level knowledge of the regular expressions could craft obfuscated reflection sequences in R, Julia, or Python to bypass string filters.

\subsection{Limitations and Threats to Validity}
\label{sec:limitations}

We note six operational limitations of the current release: (1)~\textbf{Windows dependency:} reliance on the Win32 XLL C API, COM STA threads, Named Pipes, and WebView2 restricts \neven{} to 64-bit Windows~10/11 environments; (2)~\textbf{External runtime installation:} users must install R, Julia, or Python on the host machine (guided by the automated installer), though \neven{} degrades gracefully when a subset of languages is present; (3)~\textbf{Static sandbox boundary:} as noted above, string-level filtering does not replace kernel-level mandatory access control; (4)~\textbf{Absence of formal human-factors evaluation:} while numerical accuracy is benchmarked, empirical usability gains among non-programmers remain to be quantified in a controlled classroom or industry trial; (5)~\textbf{Mocked CI host testing:} automated CI tests execute against \texttt{MockExcelBridge} rather than a live GUI Excel process; and (6)~\textbf{Bus factor:} as an academically originated open-source project, long-term sustainability requires expanding the core maintainer community beyond the primary author.

Regarding threats to validity, internal numerical validation (Section~\ref{subsec:econometric}) confirms the transparency of the IPC serialization pipeline rather than re-proving the mathematical correctness of CRAN or Julia packages. Similarly, the 2--5~ms latency figures reflect measurements on a representative desktop workstation (Intel Core i7-12700K, 32~GB RAM) and will vary with matrix dimensions and storage I/O bandwidth.


\section{Conclusions and Future Work}
\label{sec:conclusion}

\subsection{Summary and Significance}

We have presented \neven{}, an open-source C++17 polyglot infrastructure that embeds R~4.4.1, Julia~1.12.6, Python~3.13, Quarto, and interactive JavaScript visualization engines directly into Microsoft Excel. By isolating each language runtime in an independent operating system process connected via Windows Named Pipes and Protocol Buffers serialization, the platform delivers the computational rigor of modern scientific ecosystems while preserving the familiar, fault-tolerant spreadsheet interface relied upon by domain practitioners.

Through a catalog of over 70 R econometric and statistical procedures, 70 high-level Julia procedures backed by more than 200 numerical functions, zero-boilerplate user script extensibility, an 11-format WebView2 unified viewer (\texttt{=NEVEN.v()}), reactive Pluto.jl notebooks, Quarto manuscript compilation (\texttt{=NEVEN.q()}), and the nine-module NEVEN Studio IDE, \neven{} consolidates the complete analytical lifecycle within a single open-source environment. Empirical validation across the econometric benchmarks of \citet{Wooldridge2016} and dense matrix factorizations confirms that the inter-process pipeline introduces zero numerical degradation, while a suite of 342 automated unit, property-based, and security tests provides a robust foundation for academic and enterprise adoption.

\subsection{Future Work}

We identify four concrete directions for future systems and empirical research:

\paragraph{1. Kernel-Level Process Containment via Windows AppContainer}
To complement the five-layer static \texttt{SandboxVerifier}, we plan to encapsulate \texttt{ControlR.exe}, \texttt{ControlJulia.exe}, and \texttt{ControlPython.exe} within restricted Windows AppContainer security tokens and Job Objects. Enforcing mandatory access control at the OS kernel level will restrict outbound network sockets, registry modifications, and file-system writes outside designated scratch directories regardless of language-level reflection or dynamic code obfuscation.

\paragraph{2. Controlled Human-Factors Usability Study}
To empirically evaluate the pedagogical and productivity impact of the \texttt{TipoOutput} and \texttt{FILTRO} paradigms, we are designing a randomized controlled study with undergraduate econometrics and applied statistics cohorts at the Universidad de Costa Rica. The protocol will measure task completion time, specification error rates, and cognitive load (via NASA-TLX) comparing native RStudio scripting against \neven{} formula and DataLab workflows on identical econometric assignments.

\paragraph{3. Shared-Memory Zero-Copy IPC Benchmarks}
While Named Pipe serialization adds negligible latency (2--5~ms) for standard analytical tables, high-frequency quantitative finance and large-scale genomic matrices exceeding $10^6$ cells incur measurable copy overhead. We plan to benchmark a hybrid transport layer that retains Named Pipes for control framing while mapping large numeric tensors into POSIX/Win32 shared memory (\texttt{CreateFileMapping}) via Apache Arrow columnar buffers, quantifying throughput gains against PyXLL and xlwings.

\paragraph{4. Guardrails for LLM-Assisted Statistical Interpretation}
Building upon the \texttt{=P.ai\_call()} and RAG assistant infrastructure, future research will investigate structured verification guardrails that cross-check LLM-generated narrative claims against exact model diagnostics (such as $p$-value thresholds, Hausman test rejections, and variance inflation factors) extracted via \texttt{TipoOutput = N}, mitigating hallucinated statistical inferences among non-expert users.

\subsection{Availability}

\neven{} is released under the GNU General Public License v3 (GPL~v3):
\begin{itemize}
    \item \textbf{Source code and CI suite:} \url{https://github.com/minorbonillagomez/NEVEN}
    \item \textbf{Bilingual documentation:} Included in the repository (\texttt{docs/} directory, 26 chapters)
    \item \textbf{Automated Windows installer:} \texttt{Install-NEVEN.exe} (with automatic runtime detection)
    \item \textbf{Replication workbooks:} Distributed in \texttt{Ejemplos/Excel/WOOLDRIDGE/} and \texttt{Ejemplos/Excel/BONILLA/}
\end{itemize}
The platform requires 64-bit Windows~10 or~11, Microsoft Excel~2013 or later, and at least one supported runtime (R~4.4.1+, Julia~1.12.6+, or Python~3.13+).

\bibliographystyle{elsarticle-harv}
\bibliography{references}

\end{document}